\documentclass[twocolumn]{revtex4-1}

\usepackage{amsmath}
\usepackage{amsfonts}
\usepackage{graphicx}

\begin{document}

\title{Quantum Gauge Transformation, Gauge-Invariant Extension and  Angular Momentum Decomposition in Abelian Higgs Model}

\author{Israel Weimin Sun}
\email{sunwm@nju.edu.cn}

\affiliation{School of Physics, Nanjing University, Nanjing~210093, the People's Republic of China}

\date{\today}

\begin{abstract}
I discuss the momentum and angular momentum decomposition problem in the Abelian Higgs model. The usual gauge-invariant extension (GIE) construction is incorporated naturally
into the framework of quantum gauge transformation $\grave{a}$ ${\it la}$ Strocchi and Wightman and with this I investigate the momentum and angular momentum separation in a class of GIE conditions which correspond to the so-called "static gauges". Using this language I find that the so-called "generator criterion" does not generally hold even for the
dressed physical field. In the case of $U(1)$ symmetry breaking, I generalize the standard GIE construction to include the matter field degrees of freedom so that the usual
separation pattern of momentum and angular momentum in the unitarity gauge can be incorporated into the same universal framework. When the static gauge condition could not uniquely fix the gauge, I show that this GIE construction should be expanded to take into account the residual gauge symmetry. In some cases I reveal that the usual momentum or
angular momentum separation pattern in terms of the physical dressed field variables needs some type of modification due to the nontrivial commutator structure of the underlying
quantum gauge choice. Finally, I give some remarks on the general GIE constructions in connection with the possible commutator issues and modification of momentum and angular momentum separation patterns. 
\end{abstract}

\maketitle

\section{Introduction}
The study of nucleon spin structure problem has gained much progress since about 10 years before. Theoretically, the definition of the quark/gluon momentum, spin and angular
momentum operators in Quantum Chromodynamics show a multitude of possibilities which has triggered many new studies, both on the formal field theoretical side and on the actual calculational side \cite{LeaderLorce+Wakamatsu}. This new trend stems from an article of Chen and his collaborators \cite{Chen} in the year 2008. In this work the authors give a
new formulation for constructing a complete and gauge-invariant separation of the bound state spin in a gauge field system (QED/QCD). A gauge-invariant separation of the total
angular momentum operator in the QCD case was first obtained by X. Ji \cite{Ji+Chen} in 1997, however, to a large extent the separation is not unique and Chen ${\it et~al.}$'s work
paves a new way to obtain a consistent physical picture of the internal spin and momentum structure of a bound system in QED and QCD.

Soon after the appearance of the work of Chen ${\it et~al.}$ it was gradually made clear that the possible complete gauge-invariant separation of the momentum/angular momentum
operator in both QED and QCD is infinite in number and that of Chen ${\it et~al.}$ is just a specific one amongst these many possibilities. In these developments a major conceptual
progress is the use of the so-called Gauge-Invariant Extension (GIE) as a unifying language to describe the various existing momentum/angular momentum decomposition schemes.
On the formal field theoretical side, the GIE construction in fact amounts to a kind of gauge-invariant dressing of the field variables of the underlying gauge field system, or in another terminology, the so-called generalized Dirac variables. In the literature it is sometimes claimed that the whole construction program of the gauge-invariant momentum and angular momentum separations in QED/QCD can be recast into a universal formalism which uses solely the "gauge-invariant" dressed fields as the canonical variables to produce everything one wants. However, such a method is actually rather doubtful. The main reason is that the dressed fields (or the generalized Dirac variables) satisfy certain constraint
relations and could not be treated as totally independent canonical variables. As a result, their quantization method should be paid special attention, since one cannot naively
expect the standard equal-time commutation relations to hold for these "physical dressed fields".  In fact, when it is initially designed, the GIE formalism, or the dressed field, is actually a classical construction, while in the corresponding quantum gauge field theory any true discussion on the momentum/angular momentum operator decomposition issue should
be made on a specific quantum Hilbert space, which is determined by the initial quantum gauge choice. This crucial issue seems to be largely overlooked in the current literatures,
and in any sense one needs a consistent quantum field theoretical framework to understand and describe all these matters.

Such a theoretical framework actually exists, which was provided by Strocchi and Wightman in the 1970s \cite{StrocchiWightman}. In their work the authors established a general
way to implement the so-called "quantum gauge transformation" in the context of quantum electrodynamics. One could easily recognize that the GIE construction, or the physical dressed field, is actually a special case of the quantum gauge transformation $\grave{a}$ ${\it la}$ Strocchi and Wightman, when one promotes the whole classical construction
to be a quantum field theoretical construction. In terms of such a set of languages, one can make clear everything in this process, and
I believe that this language provides a suitable basis to describe the recent progress of momentum and angular momentum decomposition issues, especially in the context of nucleon spin structure studies.

The aim of this paper is to report my study of all these issues in the context of the Abelian Higgs model. This is a simple case to design and practice all the relevant theoretical
endeavors, and the existence of $U(1)$ symmetry breaking also provides some new physical insights. I show that there are some interesting phenomena in such a context and
these studies provide us more understanding on the momentum and angular momentum decomposition ways in a general gauge field theory.

\section{Basic formalism of GIE construction and decomposition of momentum and angular momentum in Abelian Higgs model}

The classical, locally $U(1)$ gauge-invariant Lagrangian density of the Abelian Higgs model has the form
\begin{equation}\label{Lagrangian1}
{\mathcal L}=-\frac{1}{4}F^2+|D_\mu \phi|^2-V(|\phi|^2)
\end{equation}
with $D_\mu\phi=(\partial_\mu+ieA_\mu)\phi$ being the covariant derivative, where $V(u)$ is a quadratic polynomial whose concrete form determines whether or not the global $U(1)$ gauge symmetry is spontaneously broken.

The standard GIE construction consists in a separation of the full gauge potential $A^\mu$ into the so-called physical component and pure-gauge component
\begin{equation}\label{phys-pure-separation}
A^\mu=A^\mu_{phys}+A^\mu_{pure}
\end{equation}
so that the $A^\mu_{phys}$ part is formally gauge-invariant while the $A^\mu_{pure}$ part transforms just like the full $A^\mu$ under an arbitrary gauge transformation. At the same
time a crucial requirement is imposed, namely, the gauge-invariant part $A^\mu_{phys}$ produces the same field strength $F^{\mu\nu}$ as $A^\mu$ does so that it actually represents
the physical degrees of freedom contained in the full gauge potential. The general defining condition for this separation is to set
\begin{equation}
f[A^\mu_{phys}]=0,
\end{equation}
which is structurally similar to a classical "gauge condition". In fact, with such a choice the "physical plus pure" separation is just a renaming of gauge potential identities.
When a specific defining condition $f[A^\mu_{phys}]=0$ is given, $A^\mu_{phys}$ is nothing but "the full gauge potential" $A^\mu$ in the special gauge choice $f[A^\mu]=0$
which has been written explicitly in terms of $F^{\mu\nu}$. This $A^\mu_{phys}$ is then called the GIE of the $A^\mu$ in a concrete form.

Now, let us recall the notion of gauge and "gauge transformation" as developed by Strocchi and Wightman in the general context of quantum electrodynamics. According to
Ref.~\cite{StrocchiWightman} this notion is first formulated for the case of a free electromagnetic field. In the general case of an indefinite metric Hilbert space, the notion of a "gauge", i.e., a quantization of the free Maxwell equations by means of a vector potential $A^\mu$, is formulated as an object $\{ A^\mu, \mathcal{H},\langle \cdot,\cdot \rangle, \mathcal{H}'\}$ which contains:
\vskip 0.2cm
\noindent (a)~A quantum field operator (or an "operator valued distribution" in the terminology of mathematicians) $A^\mu$ defined in a Hilbert space $\mathcal{H}$.
\vskip 0.2cm
\noindent (b)~A representation $U$ of the ${\rm Poincar\acute{e}}$ group in $\mathcal{H}$.
\vskip 0.2cm
\noindent (c)~A sesquilinear form $\langle \cdot,\cdot \rangle$ on $\mathcal{H}$ with respect to which the representation $U$ is unitary.
\vskip 0.2cm
\noindent (d)~A distinguished subspace $\mathcal{H}'\subset \mathcal{H}$ such that
\vskip 0.2cm
\noindent~~~~(i)~the sesquilinear form $\langle \cdot,\cdot \rangle$ is bounded and nonnegative when restricted to $\mathcal{H}'$:
\[
\langle \Psi,\Psi \rangle\geq  0,~~\forall ~\Psi \in \mathcal{H}'
\]
\noindent~~~~(ii)~the operators $F^{\mu\nu}$ (more rigorously the smeared operators $F^{\mu\nu}(f)=\int d^4x F^{\mu\nu}(x)f(x)$)
are local and leave $\mathcal{H}'$ invariant: $F^{\mu\nu}\mathcal{H}' \subset  \mathcal{H}'$,
and the free Maxwell equations hold at the level of matrix elements in $\mathcal{H}'$
\[
\langle \Phi,\partial_\mu F^{\mu\nu}\Psi \rangle=0,~~\forall ~\Phi, \Psi \in \mathcal{H}'
\]
\noindent~(iii)~the true physical state space is defined as $\mathcal{H}_{phys}=\mathcal{H}'/\mathcal{H}''$ with $\mathcal{H}''$ being the zero norm subspace of $\mathcal{H}'$,
and the representation $U$ leaves $\mathcal{H}'$ (and also $\mathcal{H}''$) invariant.
\vskip 0.2cm
\noindent~~(iv)~some additional spectral conditions which are irrelevant to our discussion.
\vskip 0.2cm
This definition has been designed to describe the usual Gupta-Bleuler quantization formalism of quantum electrodynamics. Nevertheless,
it can be easily applied to the Coulomb gauge where $\mathcal{H}=\mathcal{H}'=\mathcal{H}_{phys}$ and the Maxwell equations hold as
operator equations.

Together with such a definition of "gauge", the authors of Ref.~\cite{StrocchiWightman} also gives the following definition of "generalized gauge transformation"
\vskip 0.1cm
A generalized gauge transformation is an ordered pair consisting of two gauges
\[
\{ A^\mu_{1}, \mathcal{H}_{1},\langle \cdot,\cdot \rangle_{1}, \mathcal{H}'_{1}\} ~~{\rm and}~~\{ A^\mu_{2}, \mathcal{H}_{2},\langle \cdot,\cdot \rangle_{2}, \mathcal{H}'_{2}\}
\]
together with a bijection $g$ from $\mathcal{H}_{1 phys}$ to $\mathcal{H}_{2 phys}$ such that
\vskip 0.1cm
\noindent (i)~~$\langle \Phi_1, \mathcal{P}(F_1^{\mu\nu}(f))\Psi_1\rangle=\langle \Phi_2, \mathcal{P}(F_2^{\mu\nu}(f))\Psi_2\rangle$
\vskip 0.2cm
\noindent $\forall ~\Phi_1,\Psi_1 \in \mathcal{H}'_{1}, ~\Phi_2,\Psi_2 \in \mathcal{H}'_{2}$ with
\vskip 0.2cm
\centerline{$[\Psi_2]=g[\Psi_1],~~[\Phi_2]=g[\Phi_1]$}
\vskip 0.1cm
\noindent where $\mathcal{P}(F_{1,2}^{\mu\nu}(f))$ are polynomials in the smeared electromagnetic fields and $[\Psi_{1,2}]~(\mathrm{or}~[\Phi_{1,2}])$ denotes the equivalence class (i.e., the physical state vector) of $\Psi_{1,2}~(\mathrm{or}~\Phi_{1,2})$.
\vskip 0.2cm
\noindent (ii)~~$[\Psi_{20}]=g[\Psi_{10}]$, that is, the vacuum state maps to the vacuum state.
\vskip 0.1cm
In the general case, two different "gauges" do not necessarily share the same quantum Hilbert space $\mathcal{H}$ (together with
the associated sesquilinear form $\langle \cdot,\cdot \rangle$ and subspace $\mathcal{H}'$). However, among the generalized gauge transformations one can introduce an important subclass which is named special gauge transformations.

A special gauge transformation is a generalized gauge transformation for which the Hilbert space $\mathcal{H}$, its subspace $\mathcal{H}'$, the sesquilinear form $\langle \cdot,\cdot \rangle$, and the representation $U$ do not change and the bijection $g$ from $\mathcal{H}_{phys}$ to itself is the identity. In other words, a special gauge
transformation is just a mapping between two versions of the quantized $A^\mu$ field defined on a given Hilbert space $\mathcal{H}$ with a fixed structural input.

All the above notion of "gauge" and "gauge transformation" can be generalized to the case of interacting gauge field theories without any essential difficulty, and at the same time
with the core content retained. For such a formulation in the case of spinor quantum electrodynamics, the readers are referred to the original work of Strocchi and Wightman for more details. In this paper I will show that the usual GIE construction, or equivalently the physical dressed field, is actually a kind of "special gauge transformation" $\grave{a}$ ${\it la}$ Strocchi and Wightman, and consequently, such a theoretical framework provides a natural and appropriate language to describe all the core issues in the current discussion of momentum and angular momentum separation problems, at least in the case of quantum electrodynamics.

Now, let me describe the details of such a construction in the context of the Abelian Higgs model. First, I shall give a brief exposition at the classical level. I will use a special class of GIE conditions, which are called "static gauge" conditions, to illustrate my viewpoints.

A generic "static gauge" condition is such a linear gauge condition defined as
\begin{equation}\label{P-phys}
P^i A^i_{phys}=0,
\end{equation}
where $P^i$ is a three-vector or a three-vector of differential operators, but contains no time differentiation. One also needs to assume $P^i$ involves no explicit $x^\mu$
dependence to ensure translation invariance. To construct such an $A^\mu_{phys}$, one first notes that $A^\mu_{phys}$ should be connected with the full $A^\mu$ by a classical gauge
transformation
\begin{equation}
A^\mu_{phys}=A^\mu+\partial^\mu f,
\end{equation}
since $A^\mu_{phys}$ produces the same $F^{\mu\nu}$ as $A^\mu$ does. Then, using the condition (\ref{P-phys}) one obtains
\begin{equation}
P^i A^i_{phys}=P^i A^i+{\bf P}\cdot {\bf \partial}f=0,
\end{equation}
which yields $f=-\frac{1}{{\bf P}\cdot {\bf \partial}}P^i A^i$. This is the case when the "gauge condition"  $P^i A^i_{phys}=0$ uniquely fixes the gauge. If not, one will have
many different solutions for $f(x)$. Here, for simplicity, one assumes the gauge is uniquely fixed. I will comment on this point later.

Thus, one obtains
\begin{equation}\label{A_phys}
A^\mu_{phys}=A^\mu-\partial^\mu \frac{1}{{\bf P}\cdot {\bf \partial}}P^i A^i.
\end{equation}
At the classical level, such a construction is invariant under the gauge transformations
\[
A^\mu(x)\longrightarrow A^\mu(x)+\partial^\mu \alpha(x)
\]
with $\alpha(x)$ satisfying the boundary condition $\alpha({\bf x},t)|_{|{\bf x}|\rightarrow \infty}=0$. Such a boundary condition is to ensure that the "inverse operator"
$\frac{1}{{\bf P}\cdot {\bf \partial}}$ is uniquely defined on the function space $\{ f(x)|f= {\bf P}\cdot {\bf \partial}\alpha\}$ so that $\frac{1}{{\bf P}\cdot {\bf \partial}}
f(x)=\alpha(x)$. Mathematically, such a design of boundary condition is appropriate for the Coulomb gauge case. One assumes it also suffices at least for some class of gauge conditions $P^i A^i=0$.

Now, the gauge potential part being determined, let us look at the scalar field sector. Together with the construction (\ref{A_phys}), one establishes a dressing for the scalar
field
\begin{equation}\label{phi_phys}
\phi_{phys} = e^{ie\frac{1}{{\bf P}\cdot {\bf \partial}}P^i A^i}\phi.
\end{equation}
Then, the field variable $(A^\mu_{phys},\phi_{phys})$ is clearly invariant under the combined gauge transformations
\[
\left\{
\begin{array}{l}
 A^\mu(x)\longrightarrow A^\mu(x)+\partial^\mu \alpha(x) \\
\phi(x)\longrightarrow e^{-ie\alpha(x)}\phi(x)
\end{array}
\right.
\]
with $\alpha(x)$ satisfying the same boundary condition as before.

Of course, under a global gauge transformation
\[
\left\{
\begin{array}{l}
A^\mu(x) \longrightarrow A^\mu(x)  \\
\phi(x) \longrightarrow e^{-ie\alpha}\phi(x),
\end{array}
\right.
\]
$A^\mu_{phys}$ is unchanged whereas $\phi_{phys}$ changes by a global phase just as $\phi$
\[
\phi_{phys}(x)\longrightarrow e^{-ie\alpha}\phi_{phys}(x).
\]
This global phase degree of freedom is inherent in our construction and can be modulo away when one sticks to consider "gauge transformation" $U(x)=e^{-ie\alpha(x)}$ that tends to identity at spatial infinity.

With such a construction at hand, one can immediately find a gauge-invariant separation of the total linear and angular momentum of the coupled gauge field and scalar field system.
The recipe is very simple. The conserved momentum and angular momentum expressions derived via the Noether theorem read
\begin{equation}\label{AHM_P}
{\bf P} = \int d^3x (E^i {\bf \nabla} A^i+\pi {\bf \partial} \phi + \pi^{*}{\bf \partial} \phi^{*} )
\end{equation}
\begin{equation}\label{AHM_J}
{\bf J} = \int d^3x (E^i {\bf x}\times {\bf \nabla} A^i+{\bf E}\times {\bf A}+\pi ({\bf x}\times{\bf \partial}) \phi +\pi^{*} ({\bf x}\times{\bf \partial}) \phi^{*}),
\end{equation}
where $\pi=(D^0 \phi)^{*}=\partial_0 \phi^{*}-ie A^0\phi^{*}$ is the conjugate field.
One can equally make a dressing for the $\pi$ field variable
\begin{equation}\label{pi_phys}
\pi_{phys} = e^{-ie\frac{1}{{\bf P}\cdot {\bf \partial}}P^i A^i}\pi,
\end{equation}
resulting in a gauge-invariant construct.

Since the construction (\ref{A_phys}),(\ref{phi_phys}) and (\ref{pi_phys}) amounts to a "field-dependent gauge transformation" at the classical level and the linear and angular
momentum has a gauge-invariant appearance, one can make a direct substitution $(A^\mu,\phi,\pi)\rightarrow (A^\mu_{phys},\phi_{phys},\pi_{phys})$ in the two expressions (\ref{AHM_P}) and (\ref{AHM_J}) without changing their classical numerical values. This leads to
\begin{equation}\label{AHM_P_new}
{\bf P} = \int d^3x (E^i {\bf \nabla} A^i_{phys}+\pi_{phys} {\bf \partial} \phi_{phys} + \pi^{*}_{phys}{\bf \partial} \phi^{*}_{phys} )
\end{equation}
\begin{eqnarray}\label{AHM_J_new}
\nonumber {\bf J} &=& \int d^3x (E^i {\bf x}\times {\bf \nabla} A^i_{phys}+{\bf E}\times {\bf A}_{phys}+\pi_{phys} ({\bf x}\times{\bf \partial}) \phi_{phys} \\
 &&~~~~~~~~+\pi^{*}_{phys} ({\bf x}\times{\bf \partial}) \phi^{*}_{phys}),
\end{eqnarray}
which give a formally gauge-invariant split of the momentum and angular momentum into the contribution of their individual parts.

Now let us turn to the quantized version of such a formalism. At first sight there seems to be no essential difference between this classical construction and its quantum counterpart. This viewpoint seems to be held by most researchers working on the nucleon spin structure studies. In fact, one should remember that when quantizing a gauge field theory different quantum gauge choices usually lead to totally different quantum Hilbert spaces, and any serious discussion of the momentum or angular momentum operator decomposition problem should be built on a specific quantum gauge choice. I will show below that it is a crucial matter.

For the Abelian Higgs model there are many different quantization schemes. Among these various possibilities, one can choose, for instance, the so-called $\alpha$-gauge formalism,
which has the merit of being manifest Lorentz covariant, but as a price it leads to a Hilbert space with an indefinite metric. In our discussion I shall use the Coulomb gauge
quantization which accommodates only physical degrees of freedom. This quantization scheme can be applied to both the normal case and the $U(1)$ symmetry breaking case, although in the latter case the formal interpolating fields do not have a direct correspondence with the asymptotic fields which describe the free in/out particle states.

At a formal level, the quantum Abelian Higgs model in the Coulomb gauge is defined by the following quantum Hamiltonian operator
\begin{equation}\label{Hamiltonian}
H=\int d^3x \big[ \frac{1}{2}({\bf E}\cdot{\bf E}+{\bf B}\cdot{\bf B})+\pi^\dagger \pi+({\bf D} \phi)^\dagger \cdot({\bf D} \phi)+V(\phi^\dagger \phi) \big],
\end{equation}
where
\begin{eqnarray}
\nonumber {\bf B} &=& {\bf \nabla}\times {\bf A}_\perp \\
\nonumber {\bf E} &=& -\frac{\partial {\bf A}_\perp}{\partial t}-{\bf \nabla}A^0={\bf E}_\perp+{\bf E}_{//},
\end{eqnarray}
together with the following set of non-vanishing equal-time commutation relations (ECTRs) between its canonical variables
\begin{eqnarray}\label{commutator1}
\nonumber &&[\phi(x),\pi(y)]_{ET} = i\delta^3({\bf x}-{\bf y}) \\
\nonumber &&[\phi^\dagger(x),\pi^\dagger(y)]_{ET} = i\delta^3({\bf x}-{\bf y}) \\
 &&[A^i_\perp(x),E^j_\perp(y)]_{ET} = -i\delta^{ij}_\perp({\bf x}-{\bf y}).
\end{eqnarray}
In the Coulomb gauge quantization, $A^0$ is not an independent variable but rather satisfies the constraint
\begin{equation}\label{A0-component1}
-\nabla^2 A^0=\rho_c=ie(\phi^\dagger \pi^\dagger-\pi \phi).
\end{equation}
As a consequence, one has the following ETCRs:
\begin{eqnarray}\label{A0-phi-commutator1}
[A^0(x),\phi(y)]_{ET} &=&-\frac{e}{4\pi}\frac{\phi(y)}{|{\bf x}-{\bf y}|},
\end{eqnarray}
\begin{eqnarray}\label{A0-phi-herm-commutator1}
[A^0(x),\phi^\dagger(y)]_{ET} &=&\frac{e}{4\pi}\frac{\phi^\dagger(y)}{|{\bf x}-{\bf y}|}.
\end{eqnarray}
Therefore, the commutator $[A^0(x),\phi(y)]_{ET}$ is singular in the limit: $x \rightarrow y$. As a result, in the quantum theory the usual definition of $\pi$ field
needs to be amended. In fact, a direct use of the Heisenberg equation of motion
\[
\frac{\partial \phi(x)}{\partial t}=i[H,\phi(x)]
\]
with the quantum Hamiltonian (\ref{Hamiltonian}) yields a symmetrized expression
\begin{eqnarray}\label{pi-herm1}
\nonumber
\pi &=& \partial_0 \phi^\dagger-ie\frac{1}{2}\big(A^0\phi^\dagger+\phi^\dagger A^0\big) \\
\pi^\dagger &=& \partial_0 \phi+ie\frac{1}{2}\big(A^0\phi+\phi A^0\big).
\end{eqnarray}
These summarize the main body of the formal canonical field-theoretic inputs of the Coulomb gauge Abelian Higgs model.

Then, let us turn to the construction of momentum and angular momentum operators. It is well-known that in the Abelian Higgs model the ${\rm Poincar\acute{e}}$ symmetry is unitarily implemented in the Coulomb gauge Hilbert space which one denotes as $\mathcal{H}$
\[
U(\Lambda,a)=e^{ia_\mu P^\mu-\frac{i}{2}\omega_{\mu\nu}M^{\mu\nu}}.
\]
Among the 10 ${\rm Poincar\acute{e}}$ generators, the linear and angular momentum operators, viz. the translation and rotation generators, have the same appearance as their classical counterparts
\begin{equation}\label{AHM_P_QM}
{\bf P} = \int d^3x (E^i {\bf \nabla} A^i+\pi {\bf \partial} \phi + {\bf \partial} \phi^{\dagger} \pi^{\dagger} )
\end{equation}
\begin{equation}\label{AHM_J_QM}
{\bf J} = \int d^3x (E^i {\bf x}\times {\bf \nabla} A^i+{\bf E}\times {\bf A}+\pi ({\bf x}\times{\bf \partial}) \phi +({\bf x}\times{\bf \partial}) \phi^{\dagger} \pi^{\dagger} ),
\end{equation}
where the vector potential satisfies the Coulomb condition ${\bf \nabla}\cdot {\bf A}=0$.

Here, I emphasize that these generators are defined on the Coulomb gauge Hilbert space
$\mathcal{H}$, and in any acceptable sense one could no longer say these are "gauge-invariant" operators. In fact, in the notion of "gauge" and "gauge transformation" $\grave{a}$ ${\it la}$ Strocchi and Wightman, a particular "quantum gauge formalism" is always connected with some specific quantum Hilbert space, and what's more, one could not simply, or in
other words, naively imagine that two different "quantum gauge formalisms" are bound to be "gauge equivalent" or connected in some particular manner. Therefore, the associated Hilbert space is an indispensable element in the description of a "quantum gauge formalism", and one should not forget about this vital fact, although sometimes one might usually use the terminology "gauge invariance" to represent its classical meanings.

Then, how could one promote the classical construction (\ref{A_phys}),(\ref{phi_phys}) and (\ref{pi_phys}) to be a quantum one? The framework developed by Strocchi and Wightman
in Ref.~\cite{StrocchiWightman} provides such a natural language. In fact, based on the Coulomb gauge Hilbert space, one can promote the previous classical construction to be an operator mapping $(A^\mu, \phi,\pi)\rightarrow (A^\mu_P,\phi_P,\pi_P)$
\begin{eqnarray}\label{P_construction}
\nonumber A^\mu_P &=& A^\mu-\partial^\mu \frac{1}{{\bf P}\cdot {\bf \partial}}P^i A^i \\
\nonumber \phi_P &=& e^{ie\frac{1}{{\bf P}\cdot {\bf \partial}}P^i A^i}\phi \\
\pi_P &=& \pi e^{-ie\frac{1}{{\bf P}\cdot {\bf \partial}}P^i A^i}.
\end{eqnarray}
Such a construction is clear in meaning since no operator-ordering ambiguity arises when defining $\phi_P$ and $\pi_P$.

This operator construction is nothing but a "special gauge transformation" in the terminology of Strocchi and Wightman. The underlying Hilbert space is unique, i.e., the Coulomb gauge one, and everything is built on this unique quantum Hilbert space. Under the operator mapping (\ref{P_construction}), the field strength tensor $F^{\mu\nu}$ is unchanged,
$F^{\mu\nu}=F^{\mu\nu}_P$, and our construction ensures that the mapping $g$ of $\mathcal{H}_{phys}$ (which coincides with $\mathcal{H}$ since there exist no unphysical states in our formalism) to itself is the identity. Therefore, such a construction is a well defined one.

In the framework of Strocchi and Wightman, a special gauge transformation does not change the underlying ${\rm Poincar\acute{e}}$ group representation, and therefore the 10 ${\rm Poincar\acute{e}}$ generators $(P^\mu,M^{\mu\nu})$ will be applicable to all versions of the $(A^\mu_P,\phi_P,\pi_P)$ fields, which are defined on the same Hilbert space.
I will show below that the true meaning of a "gauge-invariant momentum and angular momentum separation" is just a reflection of the unchangeableness of the original ${\rm Poincar\acute{e}}$ generators.

The construction of the "gauge-invariant" separation of the linear and angular momentum operator is actually a process similar to the classical one, as long as one pays enough attention to the operator-ordering problems. In fact, since the ${\rm Poincar\acute{e}}$ group representation is uniquely specified by the original Coulomb gauge quantization
formalism, one can use the operator relation (\ref{P_construction}) to rewrite the translation and rotation generators (\ref{AHM_P_QM}) and (\ref{AHM_J_QM}) in terms of the newly
defined $(A^\mu_P,\phi_P,\pi_P)$ field so as to establish a new "separation pattern" for both of them. By means of the canonical formalism of the Coulomb gauge quantization, one can readily show that the relevant generators take the same form as the original operator expressions (\ref{AHM_P_QM}) and (\ref{AHM_J_QM})
\begin{equation}\label{AHM_P_QM_new}
{\bf P} = \int d^3x (E^i_P {\bf \nabla} A^i_P+\pi_P {\bf \partial} \phi_P + {\bf \partial} \phi^{\dagger}_P \pi^{\dagger}_P )
\end{equation}
\begin{equation}\label{AHM_J_QM_new}
{\bf J} = \int d^3x (E^i_P {\bf x}\times {\bf \nabla} A^i_P+{\bf E}_P\times {\bf A}_P+\pi_P ({\bf x}\times{\bf \partial}) \phi_P +({\bf x}\times{\bf \partial}) \phi^{\dagger}_P \pi^{\dagger}_P ),
\end{equation}
where the commutator structure of the Coulomb gauge quantization ensures that no other terms arise in this process. These new separation patterns show that the generators can be
 "gauge-invariantly" decomposed into contribution of the individual parts. Nevertheless, one should be cautious when speaking of such words.

The underlying reason for this is actually quite simple. In fact, in the notion of "gauge" and "gauge transformation" as formulated by Strocchi and Wightman, the field variables $(A^\mu,\phi)$ are always associated with the quantum Hilbert space $\mathcal{H}$ in which they live. When someone speaks of a "gauge transformation", he or she must go from one gauge $\langle A^\mu_1,\phi_1,\mathcal{H}_1\rangle$  to another gauge $\langle A^\mu_2,\phi_2,\mathcal{H}_2\rangle$. In this correspondence, the quantum Hilbert space usually changes, for instance, from the Coulomb gauge one to the Lorentz gauge one, and consequently one cannot naively say some construction is "gauge-invariant" in the same sense as in the classical theory, since the underlying Hilbert space has already been changed. This is also the case for the "classically gauge-invariant dressed field variables" $(A^\mu_P,\phi_P,\pi_P)$ which I have utilized in the previous constructions. Therefore, in this context one could only use the terminology "gauge-invariance" in its classical sense.

At this point, I would like to give some remarks on the generator relations concerning the dressed field $(A^\mu_P,\phi_P)$. Sometimes, it is claimed \cite{LeaderLorce+Wakamatsu} that the various parts in the total translation and rotation generators (\ref{AHM_P_QM_new}) and (\ref{AHM_J_QM_new}) act as the "translation and rotation generators" for the "physical dressed field" $(A^\mu_P,\phi_P)$. This assertion is quite misleading. In fact, one should remember that this issue needs to be checked using the commutator structure of the underlying quantum Hilbert space (which in our case is just the Coulomb gauge Hilbert space one has chosen before). Please remember that the ${\rm Poincar\acute{e}}$ generators are always unchanged, one merely changes the formulation of the relevant field degrees of freedom. One notes that the two set of operators (\ref{AHM_P_QM}) and (\ref{AHM_J_QM}) in the original Coulomb gauge act as the translation and rotation generators of the Coulomb gauge $(A^i, \phi)$ field, respectively:
\begin{eqnarray}
\nonumber [P^i,A^j(x)] &=& i \nabla^i A^j(x) \\
~[P^i,\phi(x)]&=& i\nabla^i \phi(x)
\end{eqnarray}
\begin{eqnarray}
\nonumber [J^i,A^j(x)]&=& ({\bf x}\times i{\bf \nabla})^i A^j(x)+i\epsilon^{ijk}A^k(x)  \\
~[J^i,\phi(x)]&=& ({\bf x}\times i{\bf \nabla})^i \phi(x).
\end{eqnarray}

However, this is not necessarily so for the dressed physical field $(A^i_P,\phi_P)$. In fact, one can verify immediately that under spatial translations the new $(A^i_P,\phi_P)$ field transforms in the expected manner
\begin{eqnarray}
\nonumber [P^i,A^j_P(x)] &=& i \nabla^i A^j_P(x) \\
~[P^i,\phi_P(x)]&=& i\nabla^i \phi_P(x),
\end{eqnarray}
which is actually a natural consequence of the "translation invariance" of the "GIE condition" $P^i A^i_{phys}=0$.

But the same conclusion does not hold when one speaks of the rotation transformation property of these fields. This is a logical consequence of the construction (\ref{P_construction}): $A^i$ (or $\phi$) is a three-vector (or scalar) field under spatial rotations, and $A^i_P$ (or $\phi_P$) is so only when the GIE condition $P^i A^i_{phys}=0$
is rotationally invariant, or expressed differently, the formal "3-vector" $P^i$ is a true 3-vector under the spatial rotations of the coordinate system.

Therefore, under such a situation, the "angular momentum operator" (\ref{AHM_J_QM_new}) is also the rotation generator of the physical dressed fields $(A^i_P,\phi_P)$ . But in the opposite case, one necessarily has
\begin{eqnarray}
\nonumber [J^i,A^j_P(x)]&\not=& ({\bf x}\times i{\bf \nabla})^i A^j_P(x)+i\epsilon^{ijk}A^k_P(x)  \\
~[J^i,\phi_P(x)]&\not=& ({\bf x}\times i{\bf \nabla})^i \phi_P(x).
\end{eqnarray}

This conclusion apparently disagrees with that in Ref.~\cite{LeaderLorce+Wakamatsu}. The reason is actually very simple. When the authors of \cite{LeaderLorce+Wakamatsu} made their assertions, they started from a canonical formalism in terms of the "gauge-invariant" dressed field, and in the quantized version of their theory they tacitly assumed that the "physical dressed field operators" obey the standard ETCRs of a conventional formulation of a gauge field theory, thereby arriving at the universal conclusion that the angular momentum operator should be the rotation generator of the physical dressed fields. This assumption needs not to be valid.
In fact, the underlying Coulomb gauge quantization formalism uniquely determines the commutator structure of each set of the physical dressed fields. For example, one has
\begin{equation}
[E_P^i(x),A_P^j(y)]_{ET}=i(\delta^{ij}-\frac{P^i \partial^j}{{\bf P}\cdot {\bf \partial}})\delta^3({\bf x}-{\bf y}),
\end{equation}
which deviates from the standard Coulomb gauge form and will yield a different commutator $[J^i,A^j_P(x)]$. This explains the difference between my conclusion and that in Ref.~\cite{LeaderLorce+Wakamatsu}.

It is a good lesson to compare such a kind of phenomenon with the familiar Lorentz transformation rule of the $A^\mu$ in Coulomb gauge quantization. It is a well-known fact that the gauge potential $A^\mu$ in, for instance, the free electromagnetic field theory, transforms unusually under a Lorentz boost, when one imposes the Coulomb gauge condition
${\bf \nabla} \cdot {\bf A}=0$ in every Lorentz frame. In that theory a unitary representation of the ${\rm Poincar\acute{e}}$ group is well defined in the Fock space of free transverse photons, however, the boost generator has an unusual commutator with the $A^\mu$ field, i.e., a nonstandard one which differs from that of a true four-vector field,
such that under a Lorentz boost $x'=\Lambda x$ one has \cite{BjorkenDrell}
\begin{equation}
U(\Lambda)A^\mu(x)U^{-1}(\Lambda)=(\Lambda^{-1})^\mu_{~\nu} A^\nu(x')+\frac{\partial~ \Omega(x',\Lambda)}{\partial x'_\mu},
\end{equation}
where an additional operator-valued "gauge transformation" restores the non-covariant Coulomb gauge condition ${\bf \nabla}' \cdot {\bf A}'(x')=0$ in the new Lorentz frame.

In our case, a rotationally non-covariant "gauge condition" $P^i A^i_P=0$ imposed on the $A^\mu_P$ field will render the standard commutator $[J^i,A^j_P]=({\bf x}\times i{\bf \nabla})^i A^j_P+i\epsilon^{ijk}A^k_P$ to cease to
be valid, so that even under a common spatial rotation ${\bf x}'=R{\bf x}$ one should first make a standard "rotation transformation" on the $A^i_P$ field and then supplement it by an additional gauge term
\begin{equation}
U(R)A^i_P(x)U^{-1}(R)=(R^{-1})^i_{~j} A_P^j(x')+\frac{\partial~ \Omega(x',R)}{\partial x'_i}
\end{equation}
to ensure that the "gauge condition" ${\bf P}'\cdot {\bf A}'_P(x')=0$ hold in the rotated coordinate system. This should also be the case for the dressed scalar field $\phi_P$,
where its rotation transformation rule would read
\begin{equation}
U(R)\phi_P(x)U^{-1}(R)= e^{-ie \Lambda(x',R)}\phi_P(x').
\end{equation}
Thus, one has seen that the reason of this unusual phenomenon is due to the apparent rotational non-covariance of the original GIE condition $P^i A^i_{phys}=0$, which makes our life less splendid, but in another sense more free and more flexible.

\section{Generalized GIE and unitarity gauge formalism}

In this section I shall generalize the standard GIE construction to include the scalar field degrees of freedom. This is necessary in the case of $U(1)$ symmetry being spontaneously
broken. In fact, in this case the previous translation and rotation generators, as written down in (\ref{AHM_P_QM}) and (\ref{AHM_J_QM}), could not represent a direct split into contributions of the corresponding physical particles, since the Coulomb gauge interpolating fields $({\bf A},\phi)$ do not match the true physical degrees of freedom, i.e., the asymptotic massive vector field and neutral scalar field, in a straightforward manner. In such a situation, a standard method is to make a transformation to the unitarity gauge
so that the physical degrees of freedom become transparent. I will show that this construction could also be incorporated into the framework of GIE.

As is known to all, the classical, unitarity gauge field variables are obtained from the standard $(A^\mu,\phi)$ field by means of a field-dependent gauge transformation.
One first decomposes the complex scalar field $\phi$ into the radial and angle degrees of freedom
\begin{equation}
\phi(x)=\frac{1}{\sqrt{2}}\rho(x)e^{i\theta(x)},
\end{equation}
then defines the formally gauge-invariant fields $(A^\mu_{U},\phi_{U})$ by gauging away the $\theta$ degree of freedom:
\begin{eqnarray}\label{unitarity_gauge_form}
\nonumber A^\mu_{U}&=& A^\mu+\frac{1}{e}\partial^\mu \theta  \\
\phi_{U}&=& e^{-i\theta}\phi=\frac{1}{\sqrt{2}}\rho.
\end{eqnarray}
Classically, this is a totally gauge-invariant construction, and the standard notion of GIE can be generalized so that the matter field enters into its formulation.

In the context of the Abelian Higgs model, one defines a generalized GIE to be a split: $A^\mu=A^\mu_{phys}+A^\mu_{pure}$, such that (1): the "physical part" $A^\mu_{phys}$ depends functionally on all the field degrees of freedom $(A^\mu,\phi)$; (2): it is a gauge-invariant construct under the usual gauge transformation of  $(A^\mu,\phi)$; (3): it produces the same field strength $F^{\mu\nu}$ as the full $A^\mu$ does, as a consequence, the $A^\mu_{pure}$ part has a pure gauge form.

It is apparent that the unitarity gauge variables just provide a simple example of the generalized GIE: $A^\mu=A^\mu_{phys}+A^\mu_{pure}=A^\mu_{U}-\frac{1}{e}\partial^\mu \theta$.
This generalized GIE construction differs from the previous "static gauge" GIE (\ref{A_phys}) in that it is a completely local field construction, since it just involves the fields
defined at one spacetime point. The only problem lies in the non-single-valued nature of the $\theta$ variable at the point $\rho=0$. This difficulty has a standard
way out: these types of construction should only be applied to the global $U(1)$ symmetry breaking case, where the quantum $\rho$ field operator has a nonvanishing v.e.v at the tree level $\langle \rho \rangle=\bar{\rho}$, and one only needs to consider small fluctuations of the $\rho$ field around this nonzero value.

With this GIE construction, $\phi_{U}$ can be naturally interpreted as a kind of dressing which produces a gauge-invariant and local "scalar field", and one could also introduce
a dressed $\pi$ field: $ \pi_{U}= \pi e^{i \theta}$, which is a local and gauge-invariant construct.

Now, let us turn to the momentum and angular momentum split problem.  One should remember that in the $U(1)$ symmetry breaking case it is $(A^\mu_{U},\phi_{U})$ that corresponds to
the true physical degrees of freedom, and one needs to use these variables to express everything. In the Coulomb gauge formalism, one can define a special gauge transformation by the operator mapping
\begin{eqnarray}\label{unitarity_gauge_QGT}
\nonumber A^\mu_{U} &=& A^\mu+\frac{1}{e}\partial^\mu \theta  \\
\nonumber \phi_{U}&=& e^{-i\theta}\phi=\frac{1}{\sqrt{2}}\rho  \\
\pi_{U}&=& \pi e^{i \theta}.
\end{eqnarray}
Here, it should be noted that $\phi_{U}$ can be uniquely defined since there are no operator-ordering ambiguities. However, this is not the case for $\pi_{U}$, because one has
\begin{equation}
[\pi(x),\theta(y)]_{ET}=-\frac{1}{2}\delta^3({\bf x}-{\bf y})\frac{1}{\phi(x)},
\end{equation}
which shows some intrinsic ordering ambiguity. Here, one just sticks to such a definition. Using the operator mapping (\ref{unitarity_gauge_QGT}) and paying proper attention to the operator-ordering issues, one can verify that
\begin{equation}\label{AHM_P_U}
{\bf P} = \int d^3x (E^i_U  {\bf \nabla} A^i_U+\pi_U {\bf \partial} \phi_U + {\bf \partial} \phi_U^\dagger \pi_U^\dagger),
\end{equation}
\begin{eqnarray}\label{AHM_J_U}
\nonumber {\bf J} &=& \int d^3x (E^i_U {\bf x}\times {\bf \nabla} A^i_U+{\bf E}_U \times {\bf A}_U+\pi_U ({\bf x}\times{\bf \partial}) \phi_U \\
&&~~~~~~~~+ ({\bf x}\times{\bf \partial}) \phi_U^\dagger \pi_U^\dagger).
\end{eqnarray}
To simplify all these expressions, one first notes that
\begin{equation}
\pi = \partial_0 \phi^\dagger-ie\frac{1}{2}\big(A^0\phi^\dagger+\phi^\dagger A^0\big)=\partial_0 \phi^\dagger-ieA^0\phi^\dagger+C \phi^\dagger,
\end{equation}
with $C$ being a divergent constant, then utilizing the relation (\ref{unitarity_gauge_QGT}) and the commutator structure of the Coulomb gauge quantization, one eventually obtains
\begin{equation}\label{AHM_P_U_final}
 {\bf P}=\int d^3x (E^i_U  {\bf \nabla} A^i_U+ \dot{\rho}{\bf \partial}\rho )
\end{equation}
\begin{equation}\label{AHM_J_U_final}
{\bf J} = \int d^3x (E^i_U {\bf x}\times {\bf \nabla} A^i_U+{\bf E}_U\times {\bf A}_U+ \dot{\rho}({\bf x}\times{\bf \partial})\rho ),
\end{equation}
which has the same form as the classical ${\bf P}$ and ${\bf J}$ expressions in the unitarity gauge.

The $A^i_U$ (or $\phi_U$) field defined by the operator mapping (\ref{unitarity_gauge_QGT}) is obviously a three-vector (or scalar) field under spatial rotations. Consequently,
the ${\bf P}$ and ${\bf J}$ operators are their translation and rotation generators, respectively:
\begin{eqnarray}
\nonumber [P^i,A^j_U(x)] &=& i \nabla^i A^j_U(x) \\
~[P^i,\phi_U(x)]&=& i\nabla^i \phi_U(x)
\end{eqnarray}
\begin{eqnarray}
\nonumber [J^i,A^j_U(x)]&=& ({\bf x}\times i{\bf \nabla})^i A^j_U(x)+i\epsilon^{ijk}A^k_U(x)  \\
~[J^i,\phi_U(x)]&=& ({\bf x}\times i{\bf \nabla})^i \phi_U(x).
\end{eqnarray}

The unitarity gauge GIE is a special construction. In principle there should exist many different generalized GIE constructions. At this point, I would like to make some comments on the special feature of Coulomb gauge quantization. It is true that according to the general dogma of gauge invariance, no special gauge choice is more superior to the other. However, our discussion on all the previous issues shows that the Coulomb gauge is special in at least two aspects.  First, it only contains physical transverse photons, second, this gauge choice is linear and rotational invariant, as opposed to, e.g., the general "static gauge", where explicit rotational invariance is often lost. If one insists on manifest Lorentz covariance, one may prefer to choose the Lorentz gauge, but one should necessarily meet with an indefinite metric Hilbert space which contains nonphysical degrees of freedom. When one sticks to the existence of physical degrees of freedom, the Coulomb gauge choice is indeed rather special.

\section{Some more discussions}

After all these discussions, I shall dwell on some fine points. In our previous construction of the static gauge GIE form, I assumed that
the "gauge condition" $P^i A^i_{phys}=0$ uniquely fix the gauge. In many cases this is not satisfied. A simple example is the axial gauge $A^{3}_{phys}=0$, where there is a residual
gauge symmetry: $A^\mu_{phys}\rightarrow A^\mu_{phys}+\partial^\mu f$ with $\partial^3 f=0$.

When this occurs, our previous construction for the gauge-invariant $A^\mu_{phys}$ should be amended and expanded. In fact, in such a general situation, our previous condition $P^i A^i+{\bf P}\cdot {\bf \partial}f=0$ has many solutions which can be written as
\[
f=-\frac{1}{{\bf P}\cdot {\bf \partial}}P^i A^i+f_0,
\]
where $f_0$ is an arbitrary solution of the corresponding homogeneous equation: ${\bf P}\cdot {\bf \partial} f_0=0$. In this case, a naive construction of $A^\mu_{phys}$ meets with
some difficulties, for instance, a direct choice
\[
A^\mu_{phys}=A^\mu-\partial^\mu \frac{1}{{\bf P}\cdot {\bf \partial}}P^i A^i
\]
would not be "gauge-invariant" under a residual "gauge transformation" $A^\mu \rightarrow A^\mu+\partial^\mu {\tilde \alpha}$ with ${\bf P}\cdot {\bf \partial} {\tilde \alpha}=0$.

A remedy for this actually exists. One could arbitrarily choose a specific "GIE" that is well defined, e.g., the Coulomb gauge GIE $A^\mu_C$, or in the $U(1)$ symmetry breaking case the unitarity gauge GIE $A^\mu_U$, as an initial seed of $A^\mu$ to generate everything one wants. The mthod is like this. Since the $A^\mu_{phys}$ one wants to construct is gauge-invariant and produces the full $F^{\mu\nu}$, it must differ from, for instance, the Coulomb gauge $A^\mu_C$, by a four-gradient term: $A^\mu_{phys}=A^\mu_C+\partial^\mu f$.
Then, the "gauge condition" $P^i A^i_{phys}=0$ yields $P^i A^i_C+{\bf P}\cdot {\bf \partial}f=0$, which has a general solution
\begin{equation}
f=-\frac{1}{{\bf P}\cdot {\bf \partial}}P^i A^i_C+f_0,
\end{equation}
where $f_0$ could be an arbitrary gauge-invariant construct $f_0[x:A^\mu,\phi]$ that satisfies ${\bf P}\cdot {\bf \partial} f_0(x)=0$. Thus, in this case one can define a whole family of $A^\mu_{phys}$ that satisfy the GIE condition $P^i A^i_{phys}=0$
\begin{equation}
A^\mu_{phys}=A^\mu_C-\partial^\mu \frac{1}{{\bf P}\cdot {\bf \partial}}P^i A^i_C+\partial^\mu f_0.
\end{equation}
This construction actually coincides with (\ref{A_phys}) when the condition $P^i A^i_{phys}=0$ uniquely fixes the gauge. In fact, in that case the $\partial^\mu f_0$ term vanishes
and (\ref{A_phys}) itself is a gauge-invariant construction so that one can substitute the full $A^\mu$ by the $A^\mu_C$ without changing its actual numerical values.

With this at hand, one can introduce a new type of dressing for the scalar field sector, which reads
\begin{eqnarray}
\nonumber \phi_{phys} &=& e^{ie(\frac{1}{{\bf P}\cdot {\bf \partial}}P^i A^i_C-f_0)}\phi_C \\
 \pi_{phys} &=& e^{-ie(\frac{1}{{\bf P}\cdot {\bf \partial}}P^i A^i_C-f_0)}\pi_C,
\end{eqnarray}
where $(\phi_C,\pi_C)$ is the Coulomb gauge one.

At the quantum level one can promote the above construction to be a special quantum gauge transformation on the Coulomb gauge Hilbert space. The global pattern is like this:
\begin{eqnarray}
\nonumber A^\mu_{P}&=& A^\mu_C-\partial^\mu \frac{1}{{\bf P}\cdot {\bf \partial}}P^i A^i_C+\partial^\mu f_0 \\
\nonumber \phi_{P} &=& e^{ie(\frac{1}{{\bf P}\cdot {\bf \partial}}P^i A^i_C-f_0)}\phi_C \\
\pi_{P} &=& \pi_C e^{-ie(\frac{1}{{\bf P}\cdot {\bf \partial}}P^i A^i_C-f_0)},
\end{eqnarray}
where the $f_0$ part could be any construct $f_0[x;A^\mu_C,\phi_C]$ which satisfies ${\bf P}\cdot {\bf \partial} f_0(x)=0$.

This quantum construction is a generalization of (\ref{P_construction}) where the residual gauge symmetry permits an additional term in the expression $\alpha=-\frac{1}{{\bf P}\cdot {\bf \partial}}P^i A^i_C+f_0$. The choice of the object $f_0$ is rather free and flexible, and sometimes it may render the standard separation (\ref{AHM_P_QM_new}) or (\ref{AHM_J_QM_new}) to be invalid. I will give below a concrete example to show how this comes about.

Let us consider the axial gauge case: $A^3_{phys}=0$. The construction amounts to $\alpha(x)=-\frac{1}{\partial^3} A_C^3(x)+f_0(x)$. The choice of $f_0$ is many, with the sole condition $\partial^3 f_0=0$. I shall choose a simple form:
\begin{equation}\label{Axial_gauge_construction}
f_0(x)=\frac{1}{M^2}\int_{-\infty}^{+\infty}dy^3 \phi^\dagger\phi(x^0,x^1,x^2,y^3),
\end{equation}
which is manifestly gauge-invariant and an arbitrary mass scale $M$ has been introduced for dimensional reasons.
The construction itself keeps translational symmetry and naturally satisfies $\partial^3 f_0(x)=0$.

To show that under such a condition the momentum or angular momentum separation pattern necessarily needs modification, let me analyze such an issue in some detail.
In fact, under a general operator substitution
\begin{eqnarray}\label{New}
A^\mu &\longrightarrow& A^\mu_{new}= A^\mu_C+\partial^\mu \alpha \nonumber \\
\phi &\longrightarrow& \phi_{new}=e^{-ie \alpha}\phi_C \nonumber \\
\pi &\longrightarrow& \pi_{new}=\pi_C e^{ie \alpha},
\end{eqnarray}
one has no reason to expect that the linear or angular momentum operator be kept invariant. For instance, the momentum operator (\ref{AHM_P_QM}) can be put into the form
\begin{eqnarray}\label{Momentum-new-form}
\nonumber {\bf P} &=& \int d^3x (E^i_C  {\bf \nabla} A^i_C+\pi_C {\bf \partial} \phi_C -\phi^\dagger_C{\bf \partial} \pi^\dagger_C) \\
&=& {\bf P}_{em}+{\bf P}_{scalar}.
\end{eqnarray}
Under the substitution (\ref{New}), the ${\bf P}_{em}$ part changes by an amount
\begin{equation}\label{EM-P-increasement}
\Delta {\bf P}_{em}=\int d^3x  E^i_C  {\bf \nabla}\partial^i \alpha=\int d^3x {\bf \nabla}\cdot {\bf E}_C~{\bf \nabla}\alpha,
\end{equation}
while for the ${\bf P}_{scalar}$ part, if one assumes: (1) $[\alpha(x),\partial^i \alpha(x)]=0$;~(2) $[\partial^i \alpha(x),\phi(x)]=[\partial^i \alpha(x),\pi^\dagger(x)]=0$,
one can obtain
\begin{eqnarray}
\nonumber \Delta {\bf P}_{scalar} &=& \int d^3x \big(\pi_C e^{ie \alpha} {\bf \partial} e^{-ie \alpha} ~\phi_C-\phi^\dagger_C  e^{ie \alpha} {\bf \partial} e^{-ie \alpha}~ \pi^\dagger_C)\big) \\
\nonumber &=& \int d^3x \big(\pi_C (-ie) {\bf \partial}\alpha ~\phi_C-\phi^\dagger_C  (-ie){\bf \partial}\alpha  ~\pi^\dagger_C \big) \\
\nonumber &=&  \int d^3x (-ie)(\pi_C \phi_C-\phi^\dagger_C \pi^\dagger_C){\bf \partial}\alpha \\
&=&  \int d^3x ~\rho_c {\bf \partial}\alpha.
\end{eqnarray}
Then, an application of the Gauss law ${\bf \nabla}\cdot {\bf E}_C=\rho_c$ shows $\Delta {\bf P}_{em}+\Delta {\bf P}_{scalar}=0$, so that
\begin{eqnarray}\label{P_separation_1}
{\bf P} = \int d^3x (E^i_{new}  {\bf \nabla} A^i_{new}+\pi_{new} {\bf \partial} \phi_{new} +{\bf \partial}\phi^\dagger_{new} \pi^\dagger_{new}).
\end{eqnarray}
The analysis for the angular momentum case is similar.

However, if any one of the previous assumptions concerning the commutator structure fails, one should not expect a separation like (\ref{P_separation_1}) to hold or, at the least, the separation pattern should be amended. I will show below that this is the case for our axial gauge construction (\ref{Axial_gauge_construction}).

The actual analysis is quite simple. One notes that
\begin{eqnarray}
\nonumber &&\alpha(x)=-\frac{1}{\partial^3} A_C^3(x)+f_0(x) \\
\nonumber &=& - \int_{-\infty}^{x^3}dy^3 A_C^3(x^0,x^1,x^2,y^3) \\
  &&+\frac{1}{M^2}\int_{-\infty}^{+\infty}dy^3 (\phi^\dagger \phi)_C(x^0,x^1,x^2,y^3),
\end{eqnarray}
and let me check the previous conditions one by one.

First, the condition (1) is satisfied because $[\alpha(x),\alpha(y)]_{x^0=y^0}=0$ due to the ECTRs of the Coulomb gauge. As to the condition (2), a similar commutator $[\alpha(x),\phi_C(y)]_{x^0=y^0}=0$ yields $[\partial^i \alpha,\phi_C]=0$, however, the equality $[\partial^i \alpha,\pi^\dagger_C]=0$ no longer holds, which can be seen as follows.

Let us calculate the commutator $[\alpha(x),\pi^\dagger_C(w)]_{x^0=w^0}$ using the ECTRs of the Coulomb gauge. One finds
\begin{eqnarray}
\nonumber &&[\alpha(x),\pi^\dagger_C(w)]_{x^0=w^0} = [f_0(x),\pi^\dagger_C(w)]_{x^0=w^0} \\
\nonumber &=& \frac{1}{M^2}\int_{-\infty}^{+\infty}dy^3 [(\phi^\dagger \phi)_C(x^0,x^1,x^2,y^3),\pi^\dagger_C(w)]_{x^0=w^0} \\
&=& \frac{i}{M^2} \delta(x^1-w^1)\delta(x^2-w^2)\phi_C(w),
\end{eqnarray}
thereby
\[
[\partial^i \alpha(x),\pi^\dagger_C(w)]_{x^0=w^0}=\frac{i}{M^2}
\left(
\begin{array}{c}
 \delta'(x^1-w^1)\delta(x^2-w^2)  \\
 \delta(x^1-w^1)\delta'(x^2-w^2) \\
  0
\end{array}
\right)\phi_C(w),
\]
which shows
\[
[\partial^i \alpha(x),\pi^\dagger_C(x)]=\frac{i}{M^2}\delta'(0)\delta(0)
\left(
\begin{array}{c}
  1 \\
  1 \\
  0
\end{array}
\right)\phi_C(x),
\]
with an overall singular coefficient which stems from the commutator structure of the Coulomb gauge.

With all these results, one can deduce that for this particular dressed field construction the total variation $\Delta {\bf P}_{em}+\Delta {\bf P}_{scalar}$ does not vanish and the momentum decomposition (\ref{AHM_P_QM_new}) should be modified to take into account such a correction term
\begin{eqnarray}
\nonumber {\bf P} &=& \int d^3x (E^i_P {\bf \nabla} A^i_P+\pi_P {\bf \partial} \phi_P + {\bf \partial} \phi^{\dagger}_P \pi^{\dagger}_P \\
&&+\frac{e}{M^2}\delta'(0)\delta(0)
\left(
\begin{array}{c}
 1  \\
1 \\
0\end{array}
\right)\phi^{\dagger}_P \phi_P).
\end{eqnarray}
For the angular momentum case one can find a similar modification in its separation pattern whose details will not be explicitly written out here.

\section{Some remarks on the general GIE construction}

Finally, I shall give some remarks on the general GIE construction. In this article I studied the static gauge case where its defining condition $P^i A^i_{phys}=0$
only involves the spatial components of the gauge potential $A^\mu$. This of course excludes a class of usually used GIE conditions, such as the light-front GIE condition $A^{+}_{phys}=\frac{1}{\sqrt{2}}(A^0_{phys}+A^3_{phys})=0$ which has been discussed frequently in the context of present nucleon spin structure studies. I will show that within the framework established in this article one could describe some aspects of this case.

Our previous method can be used in the light-front GIE case. In fact, by introducing a null 4-vector $P^\mu=\frac{1}{\sqrt{2}}(1,0,0,-1)$ in the light-front direction, one can write this GIE condition as $P_\mu A^\mu_{phys}=0$. Then, a direct ansatz $A^\mu_{phys}=A^\mu_C+\partial^\mu f$ will yield, for instance, $f=-\frac{1}{P\cdot \partial}P\cdot A_C=-\frac{1}{\partial^{+}}A_C^{+}$, and one writes
\begin{equation}
A^\mu_{phys}=A^\mu_C-\partial^\mu \frac{1}{\partial^{+}}A_C^{+}.
\end{equation}
One can make a dressing for the matter field sector and then promote the whole construction to be a special quantum gauge transformation in the Coulomb gauge Hilbert space:
\begin{eqnarray}
\nonumber A^\mu_P &=& A^\mu_C-\partial^\mu \frac{1}{\partial^{+}}A_C^{+} \\
\nonumber \phi_P &=& e^{ie\frac{1}{\partial^{+}}A_C^{+}}\phi_C \\
\pi_P &=& }\pi_C e^{-ie\frac{1}{\partial^{+}}A_C^{+}.
\end{eqnarray}
This is a well-defined set of quantum field variables. However, just as described in the previous section, to some extent, there seems to be no sufficient reason to believe that the translation/rotation generators could be exactly split into contributions of the various parts as in (\ref{AHM_P_QM_new}) and (\ref{AHM_J_QM_new}).

In fact, in this construction one has
\begin{equation}\label{LF}
\alpha(x)=-\frac{1}{\partial^{+}}A_C^{+}(x)=-\int_{-\infty}^0  A^{+}_C(x+\lambda n)d\lambda,
\end{equation}
where $n^\mu=\frac{1}{\sqrt{2}}(1,0,0,1)$ is another null 4-vector pointing in the light-front direction. So this $\alpha(x)$ operator construct is nonlocal in time, and this time non-locality will render the verification of the previous conditions (1) and (2) to be a nearly impossible task. For example, using (\ref{LF}) one has
\begin{equation}
[\partial^i \alpha(x),\phi(x)]=-\int_{-\infty}^0  [~\partial^i_x A^{+}_C(x+\lambda n),\phi(x)]d\lambda,
\end{equation}
which is hard to evaluate, in fact, one could not find an exact expression of the general unequal time commutator of the fundamental field operators in an interacting field theory unless one solves this theory completely.

Needless to say, the above remarks also apply to the general linear GIE condition $P_\mu A^\mu_{phys}=0$ and all these issues deserve further studies.

\section{Conclusions}

In this work I show that the GIE construction, together with the "gauge-invariant" linear and angular momentum separation issues, which is investigated in the context of the present
nucleon spin structure problem, could be naturally described by the "special quantum gauge transformation" $\grave{a}$ ${\it la}$ Strocchi and Wightman. Such a theoretical framework provides an appropriate language to formulate all these relevant notions in a gauge field system. I choose the Abelian Higgs model as a concrete example to study all these theoretical games. A special class of GIE conditions, the so-called "static gauge conditions", are used to investigate these issues. On the Coulomb gauge Hilbert space, I find a consistent picture for the quantum dressed field, translation/rotation generator decomposition and the validity of the so-called "generator criterion".

The standard GIE construction can be generalized. I show that the usual unitarity gauge field can be taken as a generalized GIE construction, based on which one can establish
the standard momentum and angular momentum separation in the $U(1)$ symmetry breaking case.

When the static gauge condition does not fix the gauge uniquely, the whole construction should be expanded to take into account the residual gauge symmetry. I find that in some cases the momentum and angular momentum separation pattern needs to be modified due to the nontrivial commutator structure of the underlying quantum Hilbert space.
A similar construction can be made for the case of a general linear GIE condition, and the same consideration suggests that the standard "gauge-invariant" split of the total momentum and angular momentum operator might need modification to some extent.

The author thanks the financial support from the Natural Science Funds of Jiangsu Province of the People's Republic of China under Grant No. BK20151376.

\end{document}